\documentclass[12pt,twoside]{article}
\usepackage{fleqn,espcrc1,epsfig}

\newcommand{\AmS}{{\protect\the\textfont2
  A\kern-.1667em\lower.5ex\hbox{M}\kern-.125emS}}

\title{\Large\bf The nuclear force in the third 
millennium\thanks{\footnotesize Invited talk 
presented at the XVIIth European Conference on Few-Body
Problems in Physics, Evora, Portugal, September 2000;
to be published in Nucl.\ Phys.\ {\bf A}.}}

\author{R. Machleidt\\ Department of Physics, University of Idaho,
        Moscow, Idaho 83844, U. S. A.}

\begin{document}

\maketitle

\begin{abstract}
I will review recent progress in our understanding of
the nuclear force. 
In the course of the 1990's, so-called high-precision,
charge-dependent
nucleon-nucleon potentials have been constructed which are, essentially,
phenomenological models. These potentials are now commonly used as
input for exact few-body calculations and microscopic nuclear many-body theory.
I will
critically analyses those models and point out their strenghts and weaknesses.
Particular emphasis will be on charge dependence. 
Other recent research was conducted on a more basic
level: understanding the nuclear force in terms of the fundamental
theory of strong interactions, QCD.
Predictions from this sector are typically qualitative in nature.
Thus, the main problem of the current status in the field 
is that quantitative models for the nuclear force
have a poor theoretical background, while theory based models
yield poor results.
The chief challenge for the new millenium is to overcome this discrepancy.
Chiral effective field theory may be a suitable tool to solve
the problem.
\end{abstract}

\section{INTRODUCTION}

The nuclear force is the heart of nuclear physics.
Yet, in spite of 70 years of research, we do not fully understand it.
I will critically review some recent advances (Sec.~2) and then
point out future directions (Sec.~3) that may have the potential
to finally crack the case---in the course of the new millennium.

\section{RECENT PROGRESS}

I define `recent progress' as the progress of the past decade
(i.~e., the 1990's). In this decade, we have seen advances
in $NN$ phenomenology as well as theory.
In this section, I will focus on phenomenology.
More fundamental issues are addressed in Sec.~3.

\subsection{Phase shift analysis}

In spite of the huge $NN$ database available today, 
conventional phase shift analyses are by no means perfect.
For example, the phase shift solutions obtained
by Bugg~\cite{BB92} or the VPI group~\cite{ASW94}
typically have a $\chi^2/$datum of about
1.4, for the energy range 0--425 MeV. 
This may be due to inconsistencies in the data
as well as deficiencies in the constraints applied in the analysis.
In any case, it is a matter of fact that
within the conventional phase shifts analysis, in which the lower partial
waves are essentially unconstrained, a better fit cannot be achieved.

About two decades ago, the Nijmegen group embarked on a program
to substantially improve $NN$ phase shift analysis.
To achieve their goal, the Nijmegen group took two 
decisive measures~\cite{Sto93}.
First, they `pruned' the data base; i.e., 
they scanned very critically the world $NN$ data base 
(all data in the energy range between zero and 350 MeV of laboratory energy 
published in a regular physics journal
between January 1955 and December 1992) and eliminated all data that
had either an improbably high $\chi^2$ 
(off by more than three standard deviations) or
an improbably low $\chi^2$; 
of the 2078
world $pp$ data below 350 MeV  
1787 survived the scan, and of the 3446 $np$ data 2514 survived.
Secondly, they introduced
sophisticated, semi-phenomenological model assumptions 
into the analysis. Namely,
for each of the lower partial waves ($J\leq 4$) a
different energy-dependent potential is adjusted
to constrain the energy-dependent analysis.
Phase shifts are obtained using these potentials in a Schroedinger
equation. From these phase shifts
the predictions for the observables are calculated including the $\chi^2$
for the fit of the experimental data. This $\chi^2$ is then minimized
as a function of the parameters of the partial-wave potentials.
Thus, strictly speaking, the Nijmegen analysis is a {\it potential analysis};
the final phase shifts are the ones
predicted by the `optimized' partial-wave
potentials which involve 39 parameters.

In the combined $pp$ and $np$ analysis~\cite{Sto93}, the fit for 
1787 $pp$ data and 2514 $np$ data below 350 MeV
results in the `perfect' 
$\chi^2/$datum = 0.99
{\it for the $NN$ database available in 1993.} 
Since then, many $pp$ data of high precision have been taken 
(particularly at IUCF~\cite{IUCF,Wis99}).
Today the $pp$ database consists of 2932 data (below 350 MeV)
for which the Nijmegen analysis produces
a $\chi^2/$datum = 1.09---not quite perfect anymore.

\subsection{The new high-precision $NN$ potentials}

In the 1990's, one focus has been on the 
quantitative aspect of the $NN$ potentials.
Even the best $NN$ models of the 1980's~\cite{Lac80,MHE87}
fit the $NN$ data typically with a $\chi^2$/datum $\approx 2$ or more.
This is still substantially above the perfect
$\chi^2$/datum $\approx 1$. 
To put microscopic nuclear structure theory to a reliable test,
one needs a perfect $NN$ potential such that discrepancies in the
predictions cannot be blamed on a bad fit of the $NN$ data.

Based upon the Nijmegen analysis and the (pruned)
Nijmegen 1992 database, new charge-dependent $NN$ potentials were
constructed in the early/mid 1990's.
The groups involved and the names of their new creations are,
in chronological order:
\begin{itemize}
\item
Nijmegen group~\cite{Sto94}: Nijm-I, Nijm-II, and Reid93 potentials.
\item
Argonne group~\cite{WSS95}: $V_{18}$ potential.
\item
Bonn group~\cite{MSS96,Mac00}: CD-Bonn potential.
\end{itemize}
All these potentials have in common that they use about 45 parameters and
fit the pruned 1992 Nijmegen data
base with a $\chi^2$/datum $\approx 1$.
However, since 1993 the $pp$ database has substantially expanded
and for the current database the $\chi^2$/datum produced by some
of these potentials is not so perfect anymore
(see Sec.~2.3.2. and Table~\ref{tab_chi2} below).

Concerning the theoretical basis of these potential, 
one could say that they are all---more or less---constructed
`in the spirit of meson theory' (e.g., all potentials include
the one-pion-exchange (OPE) contribution). However, there are
considerable differences in the details leading to considerable
off-shell differences among the potentials.

\begin{table}[t]
\caption{Modern high-precision $NN$ potentials and 
their predictions for the two- and three-nucleon bound state.}
\footnotesize
\begin{tabular}{lcccccc}
\hline
\hline
                     & 
 CD-Bonn~\cite{MSS96,Mac00}& 
 Nijm-I~\cite{Sto94}& 
 Nijm-II~\cite{Sto94}& 
 Reid93~\cite{Sto94}& 
 $V_{18}$~\cite{WSS95} &
 {\sc Nature}\\
\hline
\hline
{\it Character} & nonlocal& mixed$^a$ & local & local & local &nonlocal \\
\hline
{\it Deuteron properties:} &&&&&& \\
Quadr.\ moment (fm$^2$) & 0.270 & 0.272 & 0.271 & 0.270 & 0.270  & 
   0.276(3)$^b$ \\
Asymptotic D/S state & 0.0256& 0.0253 & 0.0252 & 0.0251 & 0.0250 & 0.0256(4) \\
D-state probab.\ (\%) & 4.85& 5.66&5.64 & 5.70  & 5.76     &  --    \\
\hline
{\it Triton binding (MeV):} &&&&&& \\
nonrel.\ calculation & 8.00 &7.72  & 7.62  & 7.63 & 7.62   & --     \\
relativ.\ calculation & 8.2  & --   &  -- & -- & -- & 8.48\\
\hline
\hline
\end{tabular}\\
$^a$ Central force nonlocal, tensor force local.\\
$^b$ Corrected for meson-exchange currents and relativity.
\label{tab_pots}
\end{table}

The CD-Bonn potential uses
the full, original, nonlocal Feynman amplitude for OPE, 
while all other potentials apply local approximations.
As a consequence of this, the CD-Bonn potential
has a weaker tensor force as compared to all other potentials.
This is reflected in the predicted D-state probabilities of
the deuteron, $P_D$, which is a measure of the strength of the nuclear tensor force.
While CD-Bonn predicts $P_D=4.85$\%, the other potentials
yield $P_D= 5.7(1)$\% (cf.\ Table~\ref{tab_pots}).
These differences in the strength of the tensor force lead to
considerable differences in nuclear structure predictions.
An indication of this is given in Table~\ref{tab_pots}:
The CD-Bonn potentials predicts 8.00 MeV for the triton binding
energy, while the local potentials predict only 7.62 MeV.
More discussion of this aspect can be found in Ref.~\cite{Mac89,Mac98}.

The OPE contribution to the nuclear force essentially takes care of the
long-range interaction and the tensor force.
In addition to this, all models must describe the intermediate
and short range interaction, for which very different
approaches are taken.
The CD-Bonn includes (besides the pion)
the vector mesons $\rho (769)$ and $\omega (783)$, 
and two scalar-isoscalar bosons, $\sigma$, 
using the full, nonlocal Feynman amplitudes for their exchanges. 
Thus, all components of the CD-Bonn are nonlocal and the off-shell
behavior is the original one as determined from
relativistic field theory.

The models Nijm-I and Nijm-II are based upon the Nijmegen78
potential~\cite{NRS78}
which is constructed from approximate one-boson-exchange (OBE) amplitudes.
Whereas Nijm-II uses the totally local approximations
for all OBE contributions, Nijm-I keeps some nonlocal
terms in the central force component (but the Nijm-I
tensor force is totally local).
However,
nonlocalities in the central force have only a very
moderate impact on nuclear structure.
Thus, it would be more important to keep the tensor force nonlocalities.

The Reid93~\cite{Sto94} and Argonne $V_{18}$~\cite{WSS95} potentials do not 
use meson-exchange for 
intermediate and short range; instead, a phenomenological parametrization
is chosen.
The Argonne $V_{18}$ uses local functions
of Woods-Saxon type, 
while Reid93 applies local Yukawa functions of multiples
of the pion mass, similar to the original Reid potential
of 1968~\cite{Rei68}.
At very short distances, the potentials are regularized 
either by 
exponential ($V_{18}$, Nijm-I, Nijm-II) or by dipole (Reid93)
form factors, which are all local functions.

\subsection{How perfect are the `perfect potentials'?}
By now, the new high-precision $NN$ potentials
have been around for a while.
It is therefore appropriate
to take a second look at them.
To properly discuss the above question,
we need to distinguish between the theoretical and practical
point of view.
On theoretical grounds, the new potentials are not perfect at all.
In fact, they are a step backwards.
Already in the 1980's, we had potentials~\cite{Lac80,MHE87}
with a better theoretical basis than any of the potentials
of the 1990's.
However, one has to concede that the new high-precision potentials
were developed essentially for practical reasons.
One wants to use them as reliable input for exact few-body 
calculations and other microscopic nuclear structure
problems. Therefore, we should judge these potentials mainly in regard
to their quantitative nature.
The question then is: Are these potentials perfect, at least,
on quantitative grounds?
We will discuss now some crucial quantitative aspects.

\subsubsection{Charge dependence}

By definition, {\it charge independence} is invariance under any 
rotation in isospin space. 
A violation of this symmetry is referred to as charge dependence
or charge independence breaking (CIB).
{\it Charge symmetry} is invariance under a rotation by 180$^0$ about the
$y$-axis in isospin space if the positive $z$-direction is associated
with the positive charge.
The violation of this symmetry is known as charge symmetry breaking (CSB).
Obviously, CSB is a special case of charge dependence.

CIB of the strong $NN$ interaction means that,
in the isospin $T=1$ state, the
proton-proton ($T_z=+1$), 
neutron-proton ($T_z=0$),
or neutron-neutron ($T_z=-1$)
interactions are (slightly) different,
after electromagnetic effects have been removed.
CSB of the $NN$ interaction refers to a difference
between proton-proton ($pp$) and neutron-neutron ($nn$)
interactions, only. 
For recent reviews, see Refs.~\cite{MO95,MNS90}.

All new potentials are charge-dependent since this is
essential for obtaining a good $\chi^2$ for the fit of the $pp$
as well as $np$ data. Thus, each potential
comes in three variants: $pp$, $np$, and $nn$.

\paragraph{Charge symmetry breaking (CSB)}

The difference between the masses of  neutron and proton
represents the most basic cause for CSB of the nuclear force. 
Therefore, it is important to have a very thorough 
accounting of this effect. 
The most trivial consequence of nucleon mass splitting is a
difference in the kinetic energies: for the heavier neutrons,
the kinetic energy is smaller than for protons. 
This raises the magnitude of the $nn$ scattering length by
0.3 fm as compare to $pp$. 
The nucleon mass difference also affects the 
one-boson-exchange (OBE) contributions to the $NN$ potential,
but only by a negligible amount. In summary, the two most obvious and trivial
CSB effects explain only about 20\% of the empirical 
CSB splitting of the $^1S_0$ scattering lenght which is
$\Delta a_{CSB} = 1.6 \pm 0.6$ fm~\cite{MNS90,How98,Gon99}.

Some models for the nuclear force (e.~g., the Nijmegen 
potentials~\cite{Sto94}) include only the two CSB effects 
just dicussed and, thus, leave CSB essentially
unexplained. 
Or, in other words, the Nijmegen group does not offer any genuine 
neutron-neutron potentials.
In some other models (e.~g., the Argonne $V_{18}$ potential~\cite{WSS95}), 
a purely phenomenological term is added
to the potential that fits $\Delta a_{CSB}$.

\begin{figure}
\begin{minipage}[t]{7.5cm}
\vspace*{-3.5cm}
\hspace*{-1.8cm}
\epsfig{figure=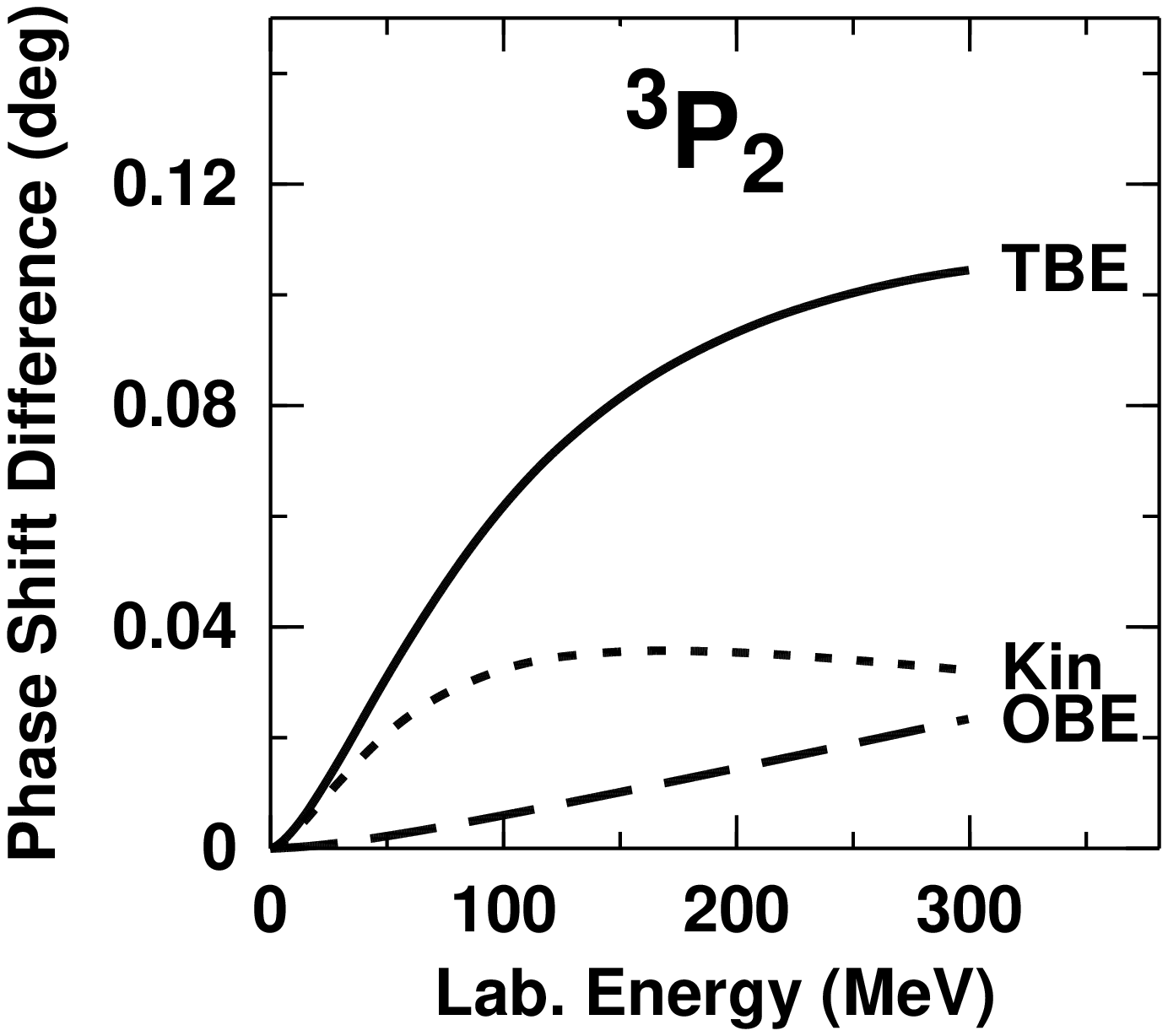,width=10.5cm}
\vspace*{-4.5cm}
\caption{CSB phase shift differences in the $^3P_2$ state due to the
impact of nucleon mass splitting on kinematics (dotted line labeled `Kin'),
one-boson exchange (dashed, OBE), and two-boson exchange diagrams
(solid, TBE).}
\label{fig_csb}
\end{minipage}
\hspace{\fill}
\begin{minipage}[t]{7.5cm}
\vspace*{-3.5cm}
\hspace*{-1.8cm}
\epsfig{figure=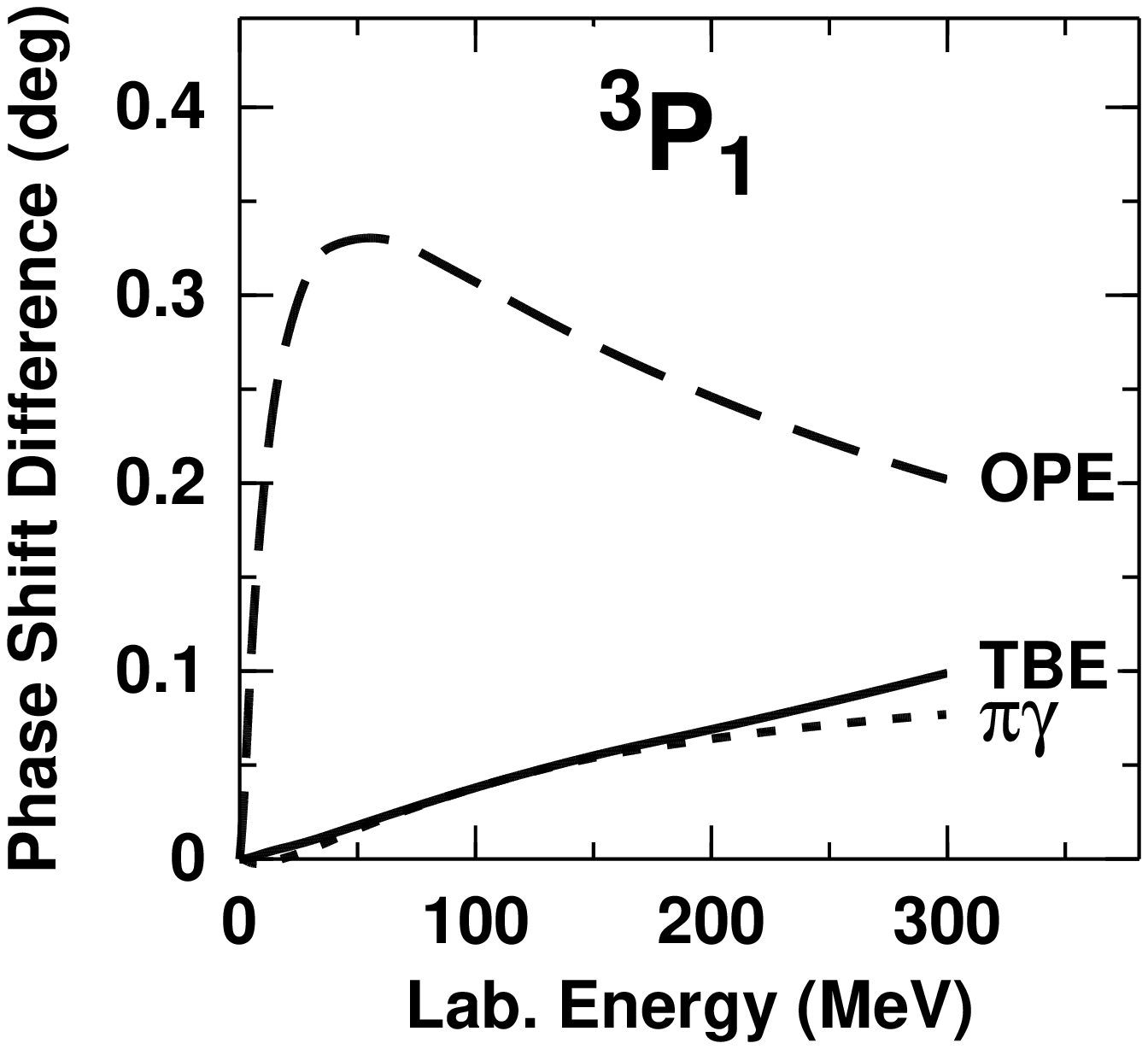,width=10.5cm}
\vspace*{-4.5cm}
\caption{CIB phase shift differences in the $^3P_1$ state 
due to the effect from pion mass splitting on OPE
(dashed line) and two-boson exchange diagrams (solid, TBE).
CIB from irreducible $\pi\gamma$ exchange 
is shown by the dotted curve.}
\label{fig_cib}
\end{minipage}
\end{figure}

The point we like to make here is that one can do much better than
this in dealing with the CSB of the nuclear force.
In Ref.~\cite{LM98a}, the CSB effect due to nucleon mass splitting
from {\it irreducible diagrams of two-boson exchange (TBE)} 
was calculated thoroughly for all two-nucleon partial waves with
$J \leq 4$.
It was found that this effect is relatively large and
fully explains
the empirical CSB splitting of the singlet scattering length.
The major part of the CSB effect comes from diagrams of
$2\pi$ exchange where those with $N\Delta$ intermediate states
make the largest contribution~\cite{CN96}. 
Noticeable CSB effects occur also in $P$ (cf.\ Fig.~\ref{fig_csb})
and $D$ waves.

Because of the outstanding importance of the CSB effect from TBE, 
it should be included in $NN$ force models 
(and, therefore, it has been incorporated
in the latest update of the CD-Bonn potential~\cite{Mac00}). 
To have distinct $pp$ and $nn$ potentials 
is important for addressing several interesting issues
in nuclear physics, like the $^3$H-$^3$He binding energy difference
for which the CD-Bonn potential predicts 60 keV 
in agreement with empirical estimates.
Another issue is the Nolen-Schiffer (NS) anomaly~\cite{NS69}
regarding the energies of neighboring mirror nuclei. 
Potentials that do not include any CSB have no chance to ever
explain this phenomenon.
Some potentials that include CSB focus on the $^1S_0$ state only, since 
this is where the most reliable empirical information is.
However, even this is not good enough.
A recent study~\cite{MPM99} has shown that
the CSB in partial waves with $L>0$
as derived from the Bonn model
is crucial for a quantitative explanation of the NS anomaly.

\paragraph{Charge independence breaking (CIB)}

The major cause of CIB in the $NN$ interaction is pion mass splitting.
Based upon the Bonn Full Model for the $NN$ interaction~\cite{MHE87},
the CIB due to pion mass splitting has been calculated 
carefully and systematically in Ref.~\cite{LM98b}.
The largest CIB effect comes from one-pion-exchange (OPE) which
accounts for about 50\% of the empirical 
CIB splitting of the $^1S_0$ scattering length, 
$\Delta a_{CIB} = 5.7 \pm 0.3$ fm~\cite{MNS90}. 
Due to the small mass of the pion, OPE is also a sizable
contribution in all partial waves with $L>0$;
and due to the pion's relatively large mass splitting (3.4\%),
OPE creates relatively large charge-dependent effects 
in all partial waves. Therefore,
all modern phase shift analyses~\cite{ASW94,Sto93} and all
modern $NN$ potentials~\cite{Sto94,WSS95,MSS96,Mac00}
include the CIB effect created by OPE.

However, pion mass splitting creates further CIB effects through
the diagrams of $2\pi$ exchange and other two-boson exchange diagrams
that involve pions. The evaluation of this
CIB contribution is very involved, but it has been accomplished
in Ref.~\cite{LM98b}. The CIB effect from all the relevant two-boson
exchanges (TBE) 
contributes about 1.3 fm to $\Delta a_{CIB}$.
Concerning phase shift differences,
it is noticeable up to $D$ waves and can amount up to 50\% 
of the OPE effect in some states
(see Fig.~\ref{fig_cib}).

Another source of CIB is irreducible $\pi\gamma$ exchange. 
Recently, these contributions have been
evaluated in the framework of chiral perturbation theory
by van Kolck {\it et al.}~\cite{Kol98}. Based upon this work,
we have calculated the impact of the $\pi\gamma$ diagrams on
the $^1S_0$ scattering length and on $np$ phase shifts.
In $L>0$ states, the size of this contribution is typically
the same as the CIB effect from TBE.

From Fig.~\ref{fig_cib} it is evident that TBE and $\pi\gamma$ create
sizable CIB effects in states with $L>0$. 
Therefore, a thoroughly constructed, modern, charge-dependent
$NN$ potential should include them.
The $NN$ potentials~\cite{Sto94,WSS95} ignore
these contributions while the latest CD-Bonn update~\cite{Mac00} 
incorporates them.

\subsubsection{After-1992 $NN$ data}

After 1992, there has been a fundamental breakthrough in 
the development of experimental methods for conducting
hadron-hadron scattering experiments.
In particular, the method of internal polarized gas targets
applied in stored, cooled beams is now working perfectly in several hadron
facilities, e.~g., IUCF and COSY.
Using this new technology, IUCF has produced a large number
of $pp$ spin correlation parameters of very high precision.
The new IUCF data~\cite{IUCF,Wis99}
(together with a few other recent $pp$ data) 
amount to 1145 pieces of new $pp$ data
(below 350 MeV).
This should be
compared to the number of $pp$ data of the pre-1993 era, namely, 1787. 
Thus, the $pp$ database has increased by about
2/3 since 1993. The importance of the new $pp$ data is further enhanced 
by the fact that they are of much higher quality than the old ones.
Therefore, the after-1992 $pp$ data represent a true challenge for
all phase shift analyses and high-quality $NN$ potentials.

In Table~\ref{tab_chi2} we show results.
The three databases used in this table are defined as follows.
The 1992 database (or pre-1993 database)
is identical to the one used by the Nijmegen group for
their phase shift analysis~\cite{Sto93}.
It consists of all $NN$ data below 350 MeV that were
published between January 1955
and December 1992 (and not rejected in the Nijmegen
data analysis). 
The after-1992 database includes all $NN$ data (below 350 MeV)
published between January 1993 and December 1999.
Finally, the 1999 database is the sum of the 1992 base 
and the after-1992 data and,
thus, represents the world $NN$ data below 350 MeV 
available in the year of 2000.

What stands out in Table~\ref{tab_chi2} are the rather large values
for the $\chi^2$/datum generated by the Nijmegen analysis and the
Argonne potential for the the after-1992 $pp$ data, which are
essentially the new IUCF $pp$ spin correlation parameters~\cite{IUCF} and
spin transfer coefficients~\cite{Wis99}. 
The Nijmegen potentials~\cite{Sto94} produce $\chi^2$ 
that are very similar to the ones by the Nijmegen analysis.
Clearly, for Nijmegen and Argonne, the $\chi^2$/datum is not
perfect anymore.
This fact is a clear indication that
these new data provide a very critical test/constraint for any $NN$
model. It further indicates that fitting the pre-1993 $pp$
data does not nessarily imply a good fit of those IUCF data. 
On the other hand, fitting the new IUCF data {\it does} imply a good fit of the
pre-1993 data. The conclusion from these two facts is that the new
IUCF data provide information that was not contained in the old database.
Or, in other words, the pre-1993 data were insufficient and still left
too much lattitude for pinning down $NN$ models.
One thing in particular that we noticed is that the $^3P_1$ phase
shifts above 100 MeV have to be lower than the values given in
the Nijmegen analysis.

\begin{table}
\caption{$\chi^2$/datum for the reproduction of the $pp$ data by
the Nijmegen phase shift analysis~\protect\cite{Sto93}, 
the Argonne $V_{18}$ potential~\protect\cite{WSS95}, and
the CD-Bonn potential~\cite{Mac00}.
The Nijmegen potentials (Nijm-I, Nijm-II, Reid93)~\protect\cite{Sto94}
produce $\chi^2$ similar to the Nijmegen PSA.}
\begin{tabular}{lccc}
\hline 
 & Nijmegen 
 & Argonne 
 & CD-Bonn 
\\

 & phase shift analysis
 & $V_{18}$ potential
 & potential
\\
\hline 
1992 $pp$ database (1787 data)   & 1.00 & 1.10 & 1.00 \\
{\bf After-1992 $pp$ data (1145 data)} & {\bf 1.24} & {\bf 1.74} & {\bf 1.03} \\
1999 $pp$ database (2932 data)   & 1.09 & 1.35 & 1.01 \\
\hline 
\end{tabular}
\label{tab_chi2}
\end{table}

\subsubsection{Extrapolating low-energy potentials to higher energies}

$NN$ potentials designed for nuclear structure purposes are typically
fitted to the $NN$ scattering data up to pion production
thereshold or slightly beyond (e.~g., 350 MeV).
A very basic reason for this is that a real potential cannot
describe the inelasticities of particle production.
On the other hand, nuclear structure calculations are
probably sensitive to the properties of a potential
above 350 MeV. For example, the Brueckner $G$-matrix,
which is a crucial quantity in many microscopic approaches
to nuclear structure, is the solution of the integral
equation,
\begin{equation}
G({\bf q'}, {\bf q}) = V({\bf q'}, {\bf q}) -
\int d^3k V({\bf q'}, {\bf k}) \frac{M^\star Q}{k^2-q^2}
G({\bf k}, {\bf q}) \; .
\end{equation}
Notice that the potential $V$ is involved in this equation for all
momenta from zero to infinity, on- and off-shell.
Now, it may very well be true that,
as the momenta increase,
their importance may decrease 
(due to the short-range repulsion of the nuclear force
and the associated
short-range suppression of the nuclear wave function).
However, it is also true that
the impact of the potential does not suddenly drop to zero
as soon as the momenta involved become larger than the
equivalent of 350 MeV lab.\ energy.
Thus, there are good physics arguments why
$NN$ potentials should
extrapolate in a reasonable way towards higher energies.

\begin{figure}
\vspace*{-4.0cm}
\hspace*{-3.0cm}
\epsfig{figure=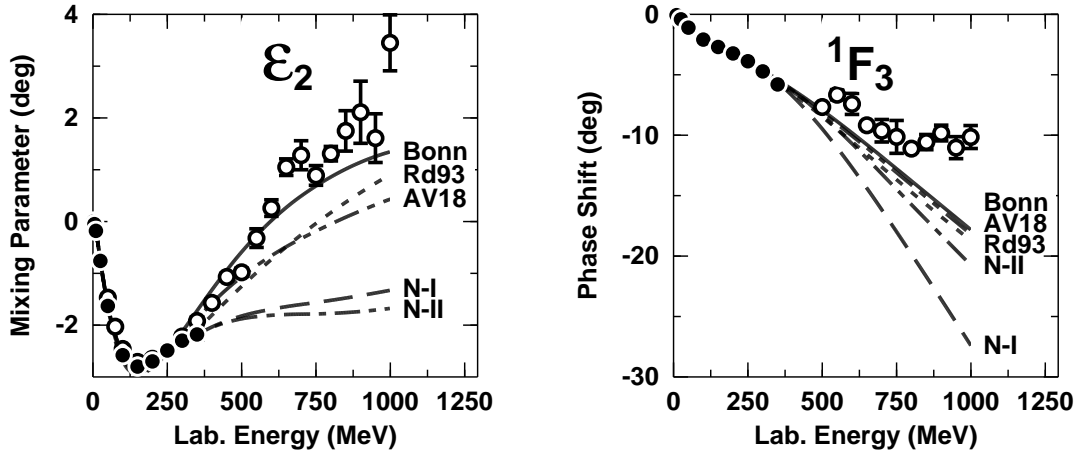,width=22.0cm}
\vspace*{-19.cm}
\caption{The $\epsilon_2$ mixing parameter and the $^1F_3$ phase shift 
up to 1000 MeV lab.\ energy for various potentials
as denoted (N-I and N-II refer to the Nijmegen potentials). 
Solid dots represent the Nijmegen PSA~\protect\cite{Sto93}
and open circles the VPI analysis SM99~\protect\cite{ASW94}.}
\label{fig_extra}
\end{figure}

We have investigated this issue and found good and bad
news. The good news is that most potentials reproduce 
in most partial waves the
$NN$ phase shifts up to about 1000 MeV amazingly well.
The bad news is that there are some singular cases
where the reproduction of phase parameters for higher energies
is disturbingly bad.
The two most notorious cases are shown in Fig.~\ref{fig_extra}.
Above 350 MeV, the $\epsilon_2$ mixing parameter is substantially
underpredicted by both Nijmegen potentials (N-I and N-II).
The reason for this is that, 
for $\epsilon_2$, 
both potentials follow 
very closely 
the Nijmegen PSA~\cite{Sto93} (solid dots in Fig.~\ref{fig_extra}) 
up to 350 MeV. 
Thus, these potentials are 
faithfull extrapolations of the Nijmegen PSA to higher energies.
Since this extrapolation is wrong, the suspicion is that
the Nijmegen PSA has a wrong trend in the energy range
250-350 MeV. New data on $pp$ spin transfer coefficients~\cite{Wis99}
in the energy range 300-500 MeV could resolve the issue.

A similar problem occurs in $^1F_3$ (Fig.~\ref{fig_extra}).
Here, the dashed curve (N-I) is the extrapolation of the
Nijmegen PSA, indicating that the analysis 
may have the wrong trend
in the energy range 200-350 MeV. 

We note that, in the two channels discussed, the inelasticity has little
impact on the phase parameters shown and would not fix the problems.

The moral is that one should not follow just one analysis,
particularly, if that analysis is severely limited in its energy range. 
It is important to also keep the broad picture in mind.

\subsubsection{Summary}

We asked the question: How perfect are the perfect 
potentials---on quantitative grounds?
The concise and discreet answer is:
Some high-quality potentials have room for improvement.

\section{FUTURE DIRECTIONS}

\subsection{Critical summary of current status}

During the past decade or so, the research on the $NN$ interaction
has proceeded essentially along two lines.
There was the phenomenological line which has produced
the high-precision, charge-dependent
$NN$ potentials~\cite{Sto94,WSS95,MSS96,Mac00}.
This was practical work, necessary to provide reliable 
input for exact few-body calculations and
nuclear many-body theory. 

The goal of the second line of research was to approach the 
problem on a more fundamental level.
Since about 1980, we have seen many efforts to derive the nuclear
force from the underlying theory of strong interactions, 
quantum chromodynamics (QCD).
Due to its nonperturbative character in the low-energy regime,
QCD cannot be solved exactly for the problem under consideration.
Therefore, so-called QCD-related or QCD-inspired models have been 
developed, like, Skyrmion or Soliton models~\cite{skyrme} 
and constituent quark cluster models~\cite{qclust,Val00}.
The success of these efforts is mixed. 
Predictions are typically only of qualitative nature;
and if predictions are quantitative then it's due to an
admixture of traditional meson exchanges.

In summary, one problem of the current status in the field 
is that quantitative models for the nuclear force
have only a poor theoretical background, while theory based models
yield only poor results.
This discrepancy between theory and practice has become rather larger 
than smaller, in the course of the 1990s.  
Another problem is that the `theory based models' are not strictly derived from
QCD, they are modeled after QCD---often with handwoven arguments.
Thus, one may argue that these models are not any better than the traditional
meson-exchange models (that are nowadays perceived as phenomenology).
The purpose of physics is to explain nature in fundamental terms.
The two trends just discussed are moving us away from this aim, which is
reason for serious concern.

Therefore, the main goal of future research on the nuclear force must be to
overcome the above discrepancies.
To achieve this goal, {\it we need a basic theory
that is amenable to calculation and yields quantitative results.}

\subsection{The Effective Field Theory Concept}

In recent years, the concept of effective field theories (EFT) 
has drawn considerable attention in particle and nuclear 
physics~\cite{Geo93,Eck95,Kap95}.
The notion of {\it effective} field theories may suggest a difference
to {\it fundamental} field theories.
However, it is quite likely that all field theories (including those
that we perceive presently as fundamental) are effective in the sense
that they are low-energy approximations to some `higher' theory.

The basis of the EFT concept is the recognition of different
energy scales in nature.
Each energy level has its characteristic degrees of freedom.
As the energy increases and smaller distances are probed, new
degrees of freedom become relevant and must be included.
Conversely, when the energy drops, some degrees of freedom 
become irrelevant and are frozen out.

To model the low-energy theory, one relies on a famous `folk theorem' by
Weinberg~\cite{Wei79,Wei97} which states:
\begin{quote}
If one writes down the most general possible Langrangian, including {\it all}
terms consistent with assumed symmetry principles,
and then calculates matrix elements with this Langrangian to any given order of
perturbation theory, the result will simply be the most general possible 
S-matrix consistent with analyticity, perturbative unitarity,
cluster decomposition, and the assumed symmetry principles.
\end{quote}
The essential point of an effective field theory is that we are not allowed to make
any assumption of simplicity about the Lagrangian and, consequently,
we are not allowed to assume renormalizability. 
The Langrangian must include all possible terms, because this completeness
guarantees that the effective theory is indeed the low-energy
limit of the underlying theory.
Now, this implies that we are faced with an infinite set of interactions.
To make the theory managable, we need to organize a perturbation expansion.
Then, up to a certain order in this expansion, the number of terms that
contribute is finite and the theory will yield a well-defined result.

In strong interactions, the transition from the `fundamental' to the
effective level happens through a phase transition that takes place
around $\Lambda_{QCD}\approx 1$ GeV 
via the spontaneous breaking of chiral symmetry 
which generates pseudoscalar Goldstone bosons.
Therefore, at low energies 
($E<\Lambda_{QCD}$), 
the relevant degrees of freedom
are not quarks and gluons, but pseudoscalar mesons and other hadrons.
Approximate chiral symmetry is reflected in the smallness of the masses
of the pseudoscalar mesons.
The effective theory that describes this scenario
is known as chiral perturbation theory ($\chi$PT)~\cite{Eck95,Leu94,BKM95}.

If we believe in the basic ideas of EFT, then, at low energies, $\chi$PT
is as fundamental as QCD at high energies.
Moreover, due to its perturbative arrangement, $\chi$PT can be calculated:
order by order. So, here we may have what we were asking for at the end
of the previous sub-section: 
{\it a basic theory that is amenable to calculation.}
Therefore, $\chi$PT has the potential to overcome the discrepancy between
theory and practice that has beset the theoretical research on the 
nuclear force for so many years.

\subsection{Chiral perturbation theory and nuclear forces}

The idea to derive nuclear forces from chiral effective Lagrangians
is not new. A program was started some 10 years ago by 
Weinberg~\cite{Wei90,Wei92}, Ord\'o\~nez~\cite{OK92},
and van Kolck~\cite{Kol93,ORK94,Kol94,Kol99}.

After the program was initiated, considerable activity 
ensued~\cite{Bea00}.
Even though all authors start from chiral effective Langrangians,
there are considerable differences in the details.
Among all these effeorts, the recent work by 
Epelbaum, Gl\"ockle, and Mei\ss ner~\cite{EGM98,EGM00,Epe00}
is particularly promising.
The details of how to derive the $NN$ potential from chiral
effective Langrangians (up to next-to-next-to-leading order)
are explained in the contribution by Epelbaum to these
proceedings~\cite{Epe00}. Therefore, I will not go into this
and, instead, recommend to the interested reader to study this beautiful
contribution.

However, I like to point out one attractive point of the
$\chi$PT approach. If, in the traditional approach,
one wants to reproduce, e.~g., the experimental binding energies of
the triton, the alpha particle or other nuclei,
one complements the $NN$ potential with a (phenomenological)
three-nucleon force (3NF)~\cite{NKG00}. Since different $NN$ potentials leave
different discrepancies to experiment (cf.\ Table~\ref{tab_pots}), 
the 3NF is adjusted
from potential to potential. 
From a more fundamental point of view, this proceedure is
very unsatisfactory, since it lacks any underlying systematics.
However, within the framework of traditional
meson theory, there is nothing else you can do,
because there is no {\it a priori} connection between the off-shell
$NN$ potential and the existence of certain many-body forces.

In the framework of $\chi$PT, there is this connection
from the outset.  In each order of $\chi$PT,
the two-nucleon force is well-defined on- and off-shell
{\it and} it is also well-defined which 3NF terms 
occur in that order.
At least that's how it should work `in theory'.
How it works out in practise remains to be seen.
We need very accurate chiral potentials to pin down these subtleties.

\section{CONCLUSIONS}

In the past decade, we had essentially two major advances.
On the practical side, high-precision phenomenological 
$NN$ potentials were constructed (some of which
need improvement to keep up with new, more accurate $NN$ data
and recent progress in charge-dependence).
On the theoretical side, nuclear physicists became
conscious of chiral symmetry (an important symmetry of QCD).
The dynamics created by this symmetry may finally
produce $NN$ potentials that are theoretically
sound {\it and} of `high-precision'.

This work was supported in part by the US
National Science Foundation under Grant-No.\ PHY-9603097.

\end{document}